\documentclass[10pt,conference]{IEEEtran}

\usepackage{cite}
\usepackage{amsmath,amssymb,amsfonts}
\usepackage{algorithmic}
\usepackage{graphicx}
\usepackage{textcomp}
\usepackage{xcolor}
\usepackage{wrapfig}
\usepackage{url}

\begin{document}

\title{Deep Learning for Low-Latency,\\Quantum-Ready RF Sensing}

\author{
    \IEEEauthorblockN{Pranav Gokhale\IEEEauthorrefmark{1},
    Caitlin Carnahan\IEEEauthorrefmark{1},
    William Clark\IEEEauthorrefmark{1},
    Teague Tomesh\IEEEauthorrefmark{1},
    Frederic T. Chong\IEEEauthorrefmark{1}\IEEEauthorrefmark{2}}
    
    \IEEEauthorblockA{\IEEEauthorrefmark{1}Infleqtion,
    Chicago, IL}
    
    \IEEEauthorblockA{\IEEEauthorrefmark{2}Department of Computer Science, University of Chicago,
    Chicago, IL}
}

\maketitle

\thispagestyle{plain} 
\pagestyle{plain} 

\begin{abstract}
Recent work has shown the promise of applying deep learning to enhance software processing of radio frequency (RF) signals. In parallel, hardware developments with quantum RF sensors based on Rydberg atoms are breaking longstanding barriers in frequency range, resolution, and sensitivity. In this paper, we describe our implementations of quantum-ready machine learning approaches for RF signal classification. Our primary objective is latency: while deep learning offers a more powerful computational paradigm, it also traditionally incurs latency overheads that hinder wider scale deployment. Our work spans three axes. (1) A novel continuous wavelet transform (CWT) based recurrent neural network (RNN) architecture that enables flexible \textit{online} classification of RF signals on-the-fly with reduced sampling time. (2) Low-latency inference techniques for both GPU and CPU that span over 100x reductions in inference time, enabling real-time operation with sub-millisecond inference. (3) Quantum-readiness validated through application of our models to physics-based simulation of Rydberg atom QRF sensors. Altogether, our work bridges towards next-generation RF sensors that use quantum technology to surpass previous physical limits, paired with latency-optimized AI/ML software that is suitable for real-time deployment.
\end{abstract}


\section{Introduction} \label{sec:introduction}

RF signal processing and sense-making are essential elements in all modern RF communications and sensing systems. In the analog domain, this processing typically includes frequency translation, filtering, and amplification, which are performed to improve signal quality for purposes of detection and demodulation. In the digital domain, after the conditioned analog signal has been sampled and digitized, the processing may include in-phase and quadrature amplitude and phase balancing, filtering, de-spreading, de-interleaving, symbol detection and synchronization, bit extraction and decoding.
The effectiveness of these methods, however, is highly dependent on the quality of the received signal, which is affected by the presence of noise, and of growing importance, by the presence of interference. As the number of emitters continues to grow, especially in urban areas, the RF spectrum is becoming increasingly congested. Moreover, in many areas of military conflict, the RF spectrum is becoming more and more contentious, with intentional interference (jamming) becoming the norm. In these environments, the received RF signal can be very dynamic and complex, with significant variations in amplitude (power), frequency, and phase content due to in-band and out-of-band interference. It is in these environments where traditional, predetermined algorithmic techniques become highly challenged, due to their lack of flexibility. AI/ML approaches, however, offer the potential for reliable real-time capability and performance despite these spectrum difficulties.

Traditionally, RF sensors have relied on conventional signal processing techniques that leverage predetermined algorithms and hard-coded system responses. While these methods are suited to handle specific types of signal interference and noise under typical conditions, they lack the flexibility to adapt to the dynamic and complex nature of real-world RF scenarios, where signal variability and unpredictable interference can drastically affect performance. AI/ML systems for signal processing offer the potential to overcome these prior limitations. Moreover, the emergence of software frameworks like PyTorch \cite{paszke2019pytorch} and CUDA/cuDNN \cite{sanders2010cuda, chetlur2014cudnn}, alongside the proliferation of GPU compute, have enabled training of deep neural networks.

There are many target applications for deep learning applied to RF signal processing including classification, anomaly detection, and novel encoding schemes. Our focus here is on classification, which has a strong basis of recent work based on the \texttt{RadioML2016.10A} dataset \cite{o2016radio}. This dataset has emerged as a standard benchmark and includes 220,000 simulated RF signals across 11 modulation modes, encompassing both digital (8 modes) and analog (3 modes) modulation types. Each signal is represented as a $2 \times 128$ array, encapsulating In-phase and Quadrature (I/Q) components across 128 time steps, with Signal-to-Noise Ratio (SNR) values ranging from -20 dB to +18 dB. The pioneering study \cite{o2016convolutional} on this dataset used a deep convolutional neural network to achieve 50.3\% test set classification accuracy (as low as 9.2\% at -18 dB SNR and 73.3\% at 10 dB SNR) after 100 epochs on training \cite{radioml_github}. The inference time was approximately 200 ms per batch of 1024, using NVIDIA GeForce GTX Titan X hardware; note that GPU performance has increased significantly in the past 8 years.

Follow-up work on the \texttt{RadioML2016.10A} dataset has led to steady progression in performance, with classification accuracy as high as 92.3\% achieved in high-SNR settings. \cite{tekbiyik2020robust}. Other work has focused on balancing model size with performance, leading to lightweight models (PET-CGDNN) that use 8x fewer parameters than competing models, while managing to maintain classification accuracy at high-SNRs above 90\% \cite{zhang2021efficient}. PET-CGDNN achieved inference time of 5.0 ms per batch of 128 samples. Yet more recent work has achieved highest classification accuracy of 98.5\% and average classification accuracy of 68.3\%, by using deeper convolutional neural networks \cite{wang2022survey}.

This paper aims to extend the state of the art in two ways. First, we seek for further optimize inference latency, so that deep learning methods can be suitable for real-time deployment. Second, we seek to apply deep learning methods to the output of emerging Quantum RF sensors---specifically ones based on Rydberg atoms. The remainder of this paper is structured as follows. Section~\ref{sec:online_classification} introduces a novel model that can perform \textit{online} classification on-the-fly of RF signals with minimal sampling time, enabling real-time processing. Section~\ref{sec:latency_optimization} studies latency optimization during inference time for both GPU and CPU, uncovering over 100x of potential gains and ultimately leading to sub-millisecond runtime. Finally, Section~\ref{sec:quantum_rf_signal_processing} applies our latency-optimized models to the simulated output of a Rydberg atom QRF sensor that simultaneously detects multiple RF tones. Section \ref{sec:conclusion} concludes and suggests directions for future work.

\section{Online Classification with Wavelet Transforms and Recurrent Networks} \label{sec:online_classification}

Radio frequency signal analysis is commonly performed in both the time and frequency domains, to extract features that can be used to improve detection and classification of received signals. Transformation from the time domain to the frequency domain is typically done using the Fast Fourier Transform (FFT) method, which reveals frequency-dependent features not readily obvious in the conjugate time domain. These features serve as a natural and effective way to train neural network-based models for RF signal analysis \cite{MoralesFerre2019, Elyousseph2021}. 

However, the standard Fourier transform poses a challenge for the task of low-latency RF classification and, indeed, any low-latency RF analysis task. Precise transformation to the frequency domain requires a long sampling time over the original signal. Furthermore, the transformed representation loses all time resolution in the signal. Such a representation is suitable for long time samples of static waveforms, but RF signals of interest fluctuate rapidly on short timescales and generally exhibit highly dynamic characteristics. Thus, frequency-domain transformations of RF signals should aim to convey the temporal dependence of the frequency spectrum.

Arbitrary precision in both time and frequency, however, is not possible due to the Gabor limit, which specifies an uncertainty principle in time and frequency; in other words, higher precision in the time domain yields lower precision in the frequency domain and vice versa. There exist several techniques---such as the Short Time Fourier Transform (STFT) and the Wavelet Transform---for acquiring an approximate time-resolved spectrum. STFTs have been used successfully in \cite{MoralesFerre2019}  to train a convolutional ML model to identify jamming in a full RF spectrogram. In Section~\ref{sec:wavelet_transform}, we will discuss the Wavelet Transform  and motivate its use in support of low-latency RF signal analysis.  

With an approximate time-resolved spectrum in hand, one may construct a neural network suitable for the given RF signal analysis task in question. In the cases of modulation and jamming classification, recent work has shown convolutional neural networks over both the time- and frequency- domain signals to be effective \cite{o2016convolutional,wang2022survey, MoralesFerre2019, West2017}. However, the convolutional layers in these models are trained to detect and classify according to fixed sample sizes of the RF signal. Such a model is at odds with the imperative of low-latency real-time analysis, in which one aims to analyze the signal within a minimal number of timesteps (i.e. the shortest signal duration possible), or to potentially incorporate additional streaming signal input as needed for high confidence output. To that end, in Section~\ref{sec:rnn}, we investigate the use of a simple recurrent neural network (RNN) designed to ingest the wavelet transformed signal incrementally by timestep and produce an \textit{online} classification result. We show that a simple model achieves impressive classification accuracy over both modulation and SNR classification tasks and, crucially, is able to maximize its prediction accuracy after ingesting a small fraction of the signal. These results indicate that RNN-based models operating on wavelet transforms may be able to support accurate and rapid real-time decision-making on incoming RF signals. Furthermore, RNN-like units may serve as the foundation for autoencoder-style models capable of more sophisticated RF signal tasks, such as signal cleaning and prediction. 

\subsection{The Continuous Wavelet Transform (CWT)} \label{sec:wavelet_transform}
Fourier-related transforms that attempt to balance time and frequency resolution include the Short Time Frequency Transform (STFT) and the Wavelet Transform. The STFT simply prescribes performing the Fourier transform on individual fixed-length short time windows over the original time-domain signal. In doing so, one can obtain an approximate time-resolved spectrum. The wavelet transform, on the other hand, convolves the original signal with a plane wave modulated by a Gaussian envelope,  
\begin{equation}
w = e^{2i\pi f t}e^{-t^2/2\sigma^2}.
\end{equation}
A novel trait of the wavelet transform is that the Gaussian envelope width, characterized by $\sigma$, provides the time localization as a function of the frequency $f$ and can be considered a tunable hyperparameter in RF signal analysis. For $\sigma \sim n/f$, one may fix $\sigma$ itself or one may fix $n$, the number of cycles, allowing the envelope to dilate and contract with the frequency. More sophisticated frequency dependence rules may be enforced to optimize spectral input for model understanding.  

\begin{figure*}[h]
    \centering
    \includegraphics[width=1.0\linewidth]{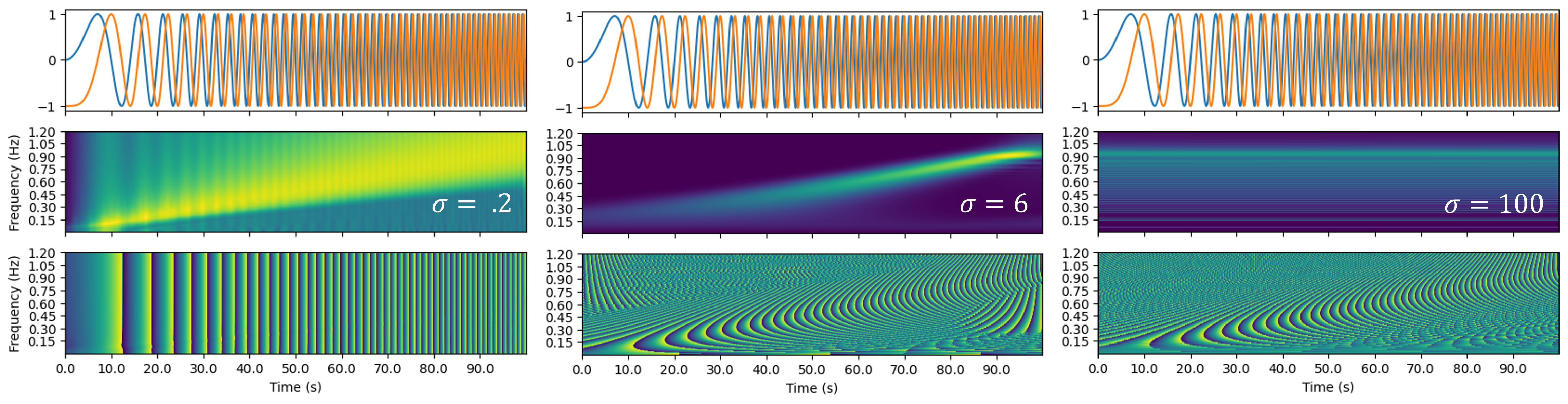}
    \caption{Left to right: plots of $S(t)$ (top), amplitude $A(t,f)$ spectrogram (middle), and phase $\phi(t,f)$ spectrogram (bottom) for Gaussian standard deviation $\sigma = .2, 6, 100$. }
    \label{fig:wavelets}
\end{figure*}

To understand the general properties of the wavelet transform, we consider the time-dependent signal  
\begin{equation}
S(t) = \sin(\pi (.1 + t^2/100)) - i\cos(\pi (.1 + t^2/100))
\end{equation}
plotted in Fig. \ref{fig:wavelets} with both real and imaginary components. The complex wavelet transform is performed with a Morlet wavelet, a complex exponential carrier modulated by a Gaussian window (in this complex form, it is also known as the Gabor wavelet). To perform the tranformation, we utilized the fast Continuous Wavelet Transform (fCWT) package which provides a highly-optimized implementation of the continuous wavelet transform \cite{Arts2022}.  The transform is performed with three different values of $\sigma$, demonstrating the trade-off in time and frequency resolution. We plot the amplitude $A$ of the frequency-domain components as well as the phase $\phi$. For small $\sigma$, the amplitude and phase spectrograms exhibit precise time resolution, but unclear localization along the frequency axis. On the other hand, for the largest $\sigma$ \textemdash approaching the limit of the standard Fourier Transform \textemdash the spectrograms exhibit precise frequency localization, but lose understanding of time dependence. For a careful choice of $\sigma$, optimized for a specific analysis task and relevant RF characteristic features, balanced precision in the two regimes may be obtained simultaneously.

Throughout the remainder of this section, we consider wavelet transformats performed over the I/Q signals in the \texttt{RadioML2016.10A} dataset. We reinterpret the dataset in units of seconds, such that each example constitutes $2\times 128$ data points taken over $128$ s at a rate of $1$ Hz. We use the nominal choice of $\sigma = 2$ and sample $32$ frequency values between $1/128$ Hz and $.3$ Hz to generate the spectrograms of the amplitude $A(t,f)$ and phase $\phi(t,f)$. Therefore, at any given time $t'$, we have a $64$-element vector $(A(t'), \phi(t'))$ that is used as input to our model, with $A(t')$ and $\phi(t')$ each composed of $32$ frequency-dependent elements. We make a special note about this observation: the wavelet transform has increased the number of parameters that we use to specify the input at any given time (namely, from two values, $I(t)$ and $Q(t)$, to $64$ values). In principle, not much more information is present, aside from limited knowledge about neighboring points used to construct the localized frequency information. However, the goal is to create a representation of the information from which patterns are more easily deduced by the model, which we demonstrate in the following section.  

\begin{figure}[h]
    \centering
    \includegraphics[width=.9\linewidth]{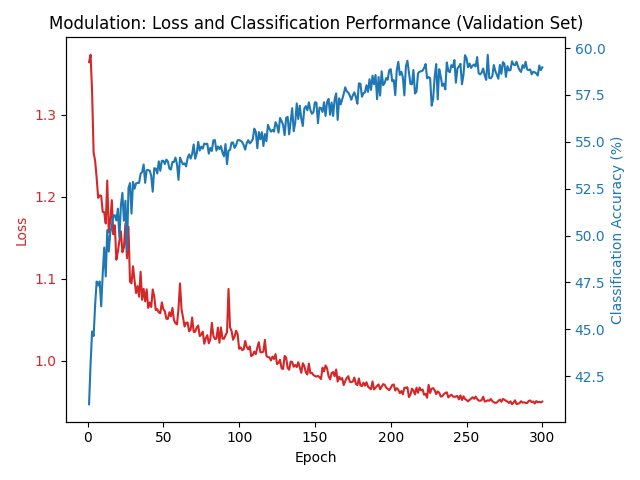}
    \caption{Validation loss and classification accuracy as a function of training epoch for the 11-modulation classification task, using our CWT-RNN model. Final classification accuracy over the validation set approached $60\%$ overall. }
    \label{fig:mod_perf}
\end{figure}

\subsection{RNN-Based Classifiers} \label{sec:rnn}
We aim to construct a model that facilitates rapid real-time RF signal analysis. As data streams in and is transformed via the wavelet transforms described in the preceding section, we obtain a time series of $64$-element input vectors $(A(t), \phi(t))$. Time-series analysis via machine learning is commonly tackled with recurrent neural networks. While the exact architecture may vary widely, a common characteristic of RNNs is that they ingest the data one timestep at a time and maintain a so-called hidden state $h_t$, which is modified during every inference step and used as input in the subsequent inference step. In this way, $h_t$ may accumulate historic information about the time series, providing a mechanism for preserving context. 

For this discussion, we construct a relatively simple RNN of one recurrent unit containing 4 linear layers and one hidden state of 64--128 parameters. Our model additionally features two LeakyReLU \cite{maas2013rectifier} activations, and Layer Normalization \cite{ba2016layer} is executed over the amplitude inputs. To enable classification over $N$-classes, a LogSoftmax operation is applied to the $N$-element output layer, over which the Negative Log Likelihood loss is calculated. During training, we use an AdamW optimizer \cite{loshchilov2017decoupled} and a learning rate scheduler featuring Linear or Exponential rate decay. Training is performed over 100--300 epochs with batch sizes of 512--1024 signals, and dropout is added to the first layer with probability 20\%--50\%. We refer to this model as the CWT-RNN model.

\begin{figure*}[h]
    \centering
    \includegraphics[width=0.8\linewidth]{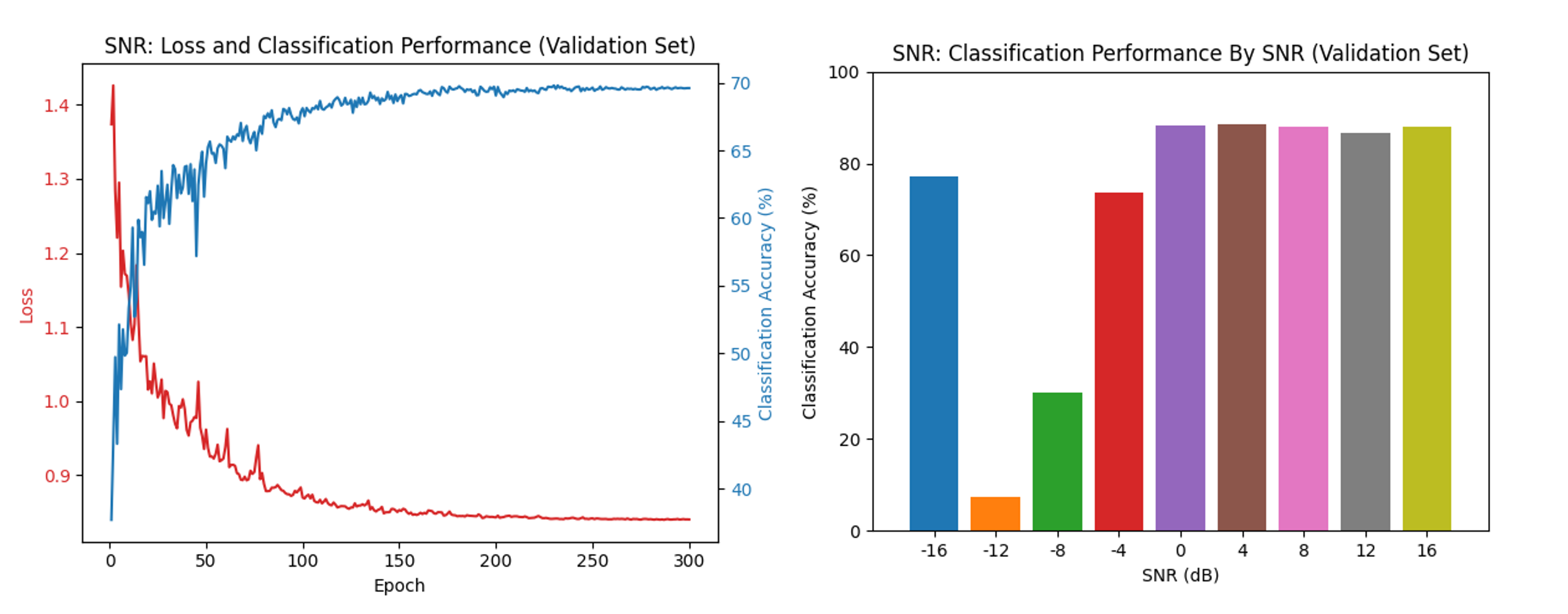}
    \caption{CWT-RNN validation loss and classification accuracy as a function of training epoch for the 9-SNR classification task. Final classification accuracy over the validation set approached $70\%$ overall. }
    \label{fig:snr9}
\end{figure*}
\begin{figure*}[h]
    \centering
    \includegraphics[width=0.8\linewidth]{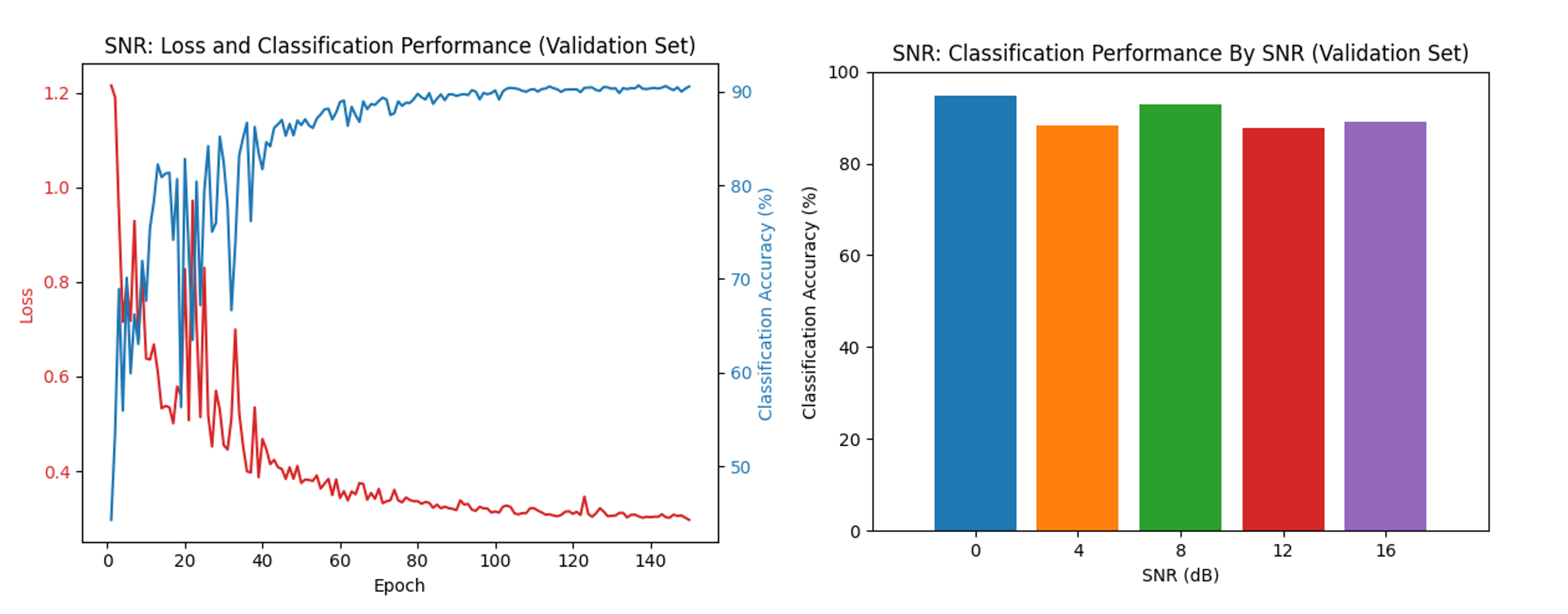}
    \caption{CWT-RNN validation loss and classification accuracy as a function of training epoch for the 5-SNR classification task. Final classification accuracy over the validation set reached $90\%$ overall. }
    \label{fig:snr5}
\end{figure*}

To draw a parallel with previous work, we first consider the performance on the modulation classification task considered in \cite{o2016convolutional}. We train and evaluate our CWT-RNN model on a high-SNR subset (0--18 dB) containing 84,000 training examples and 25,000 validation examples. The loss and classification accuracy as a function of the training epoch are shown in Fig. \ref{fig:mod_perf}. We note that the definition of classification accuracy used in this section is akin to recall: the percentage of top-1 true positives in all examples considered. We see that averaged across all SNR levels, the classification accuracy is approximately $60\%$. In reality, the per-modulation performance exhibits a bimodal distribution with 5 modulations demonstrating classification accuracy of 87\% or above. While the modulation classification of our model does not meet the performance of achieved by recent results like PET-CGDNN \cite{wang2022survey}, we emphasize that our model is relatively small in size with limited complexity. For this reason, the model size may need to be increased to effectively capture the decision boundaries between the remaining classes. Nevertheless, we note that an overall accuracy of 60\% is a significant improvement over the 9\% baseline for the 11-modulation classifier, and therefore demonstrates the ability of the CWT-based RNN to effectively learn the critical features for modulation classification.

Motivated by the critical task of perceiving RF quality and detecting interference (whether by significant noise, environmental influence or intentional jamming), we consider the alternative task of classifying signals by their SNR. We considered a task of SNR classification over the \texttt{RadioML2016.10A} dataset for $9$ SNR classes spanning $-16$ dB to $16$ dB, as well as a reduced task of categorizing the high-SNR split of $0$ dB to $16$ dB. We train and evaluate our model on $80,000$ training examples and $18,000$ validation examples for the $9$-SNR classification task, and $44,000$ training examples and $10,000$ validation examples for the $5$-SNR classification task.The loss and classification accuracy over the validation sets are shown in Figs. \ref{fig:snr9} and \ref{fig:snr5}. We find that the $9$-SNR classification task achieves 70\% accuracy on the validation set, with higher performance in the high-SNR regime. This is further confirmed by the $5$-SNR classification results, which obtain 90\% accuracy overall. The increased complexity needed to identify the decision boundaries for the low-SNR regime, as with the difficult modulation classes, can be addressed by increasing the model size or by considering alternative RNN models with more sophisticated contextuality mechanisms, such as the Long Short Term Memory (LSTM) Model.  
\begin{figure}[h]
    \centering
    \includegraphics[width=\linewidth]{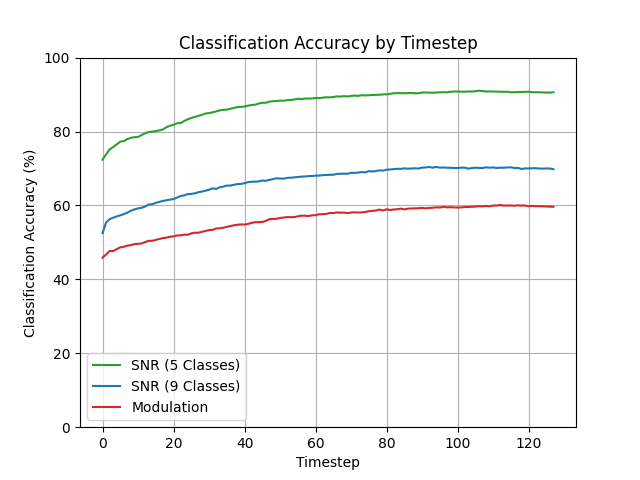}
    \caption{CWT-RNN classification accuracy as a function of timestep, averaged over the validation dataset, for all three classification tasks: Modulation, 9-SNR, and 5-SNR. }
    \label{fig:per_timestep}
\end{figure}
 
 We assert that these results demonstrate the capacity for CWT-RNNs to perform highly on relevant RF analysis tasks, including both modulation scheme detection as well as signal quality assessment. However, the pivotal advantage of the CWT-RNN model is to support real-time low-latency RF decision-making. In Fig. \ref{fig:per_timestep}, we show the average classification accuracy of the CWT-RNN model as a function of timestep for each task considered. As time-series inputs are fed into the model in an online fashion, the output for each timestep may be evaluated as a preliminary classification, the average accuracy of which is shown here. The CWT-RNN exhibits remarkably high accuracy from the first input, with the maximum accuracy achieved at only a fraction of the signal length. These results indicate that the CWT-RNN approach can not only support highly-accurate RF signal analysis, but is capable of rapid online decision-making in real time with minimal signal input. Furthermore, we emphasize that the CWT-RNN facilitates a \emph{flexible} approach where additional time-series inputs may be continuously provided to the model as needed to improve the confidence of the output.   

\section{Latency Optimization for Inference} \label{sec:latency_optimization}

A potential disadvantage of deep learning methods, relative to conventional methods for signal processing, is latency. Fielded RF systems generally have stringent latency needs, though exact requirements vary depending on the target application. For instance, radar applications generally tolerate millisecond latencies, whereas some electronic warfare systems require nanosecond response times \cite{iriarte2017lowlatency}.

\begin{figure*}[h]
    \centering    \includegraphics[width=0.8\linewidth]{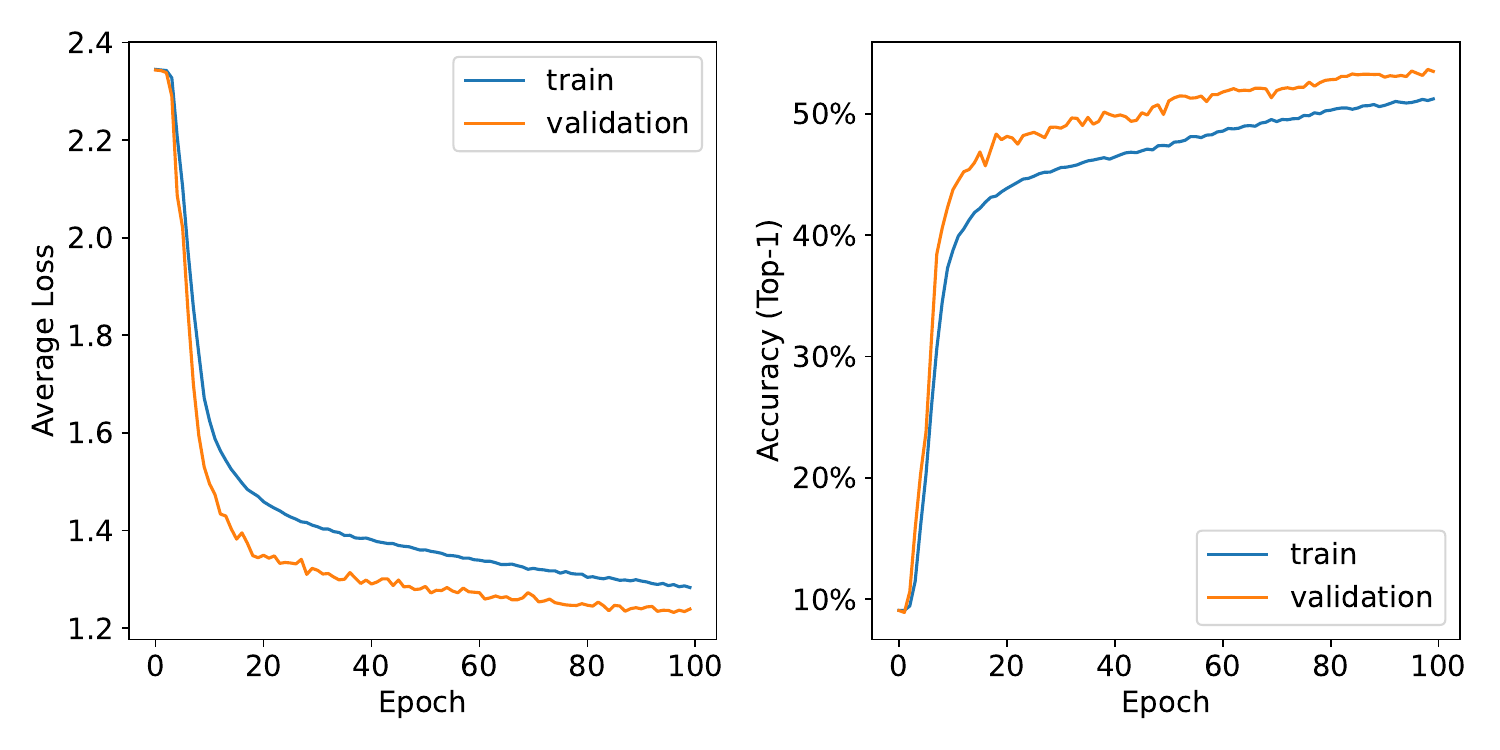}
    \caption{Loss (left) and classification accuracy on the \texttt{RadioML2016.10A} dataset using the CNN2 deep convolutional neural network \cite{o2016convolutional}, trained over 100 epochs. Training was performed with a batch size of 1024 and the Adam \cite{kingma2014adam} optimizer. During these 100 epochs, performance on the validation set is slightly better; this is an artifact of the network having Dropout layers that are disabled in evaluation model.}
    \label{fig:Loss_and_Accuracy}
\end{figure*}

One avenue for improving latency is to change the underlying deep learning model, as has been explored in prior work \cite{zhang2021efficient, wang2022survey}. Here, our focus is on optimizations that are agnostic to model architecture and instead apply across virtually every deep learning model. Nonetheless, to make our benchmarking analysis concrete, we choose---as a reference model---the \textit{CNN2} model described in \cite{o2016convolutional}. We implemented this convolutional neural network in PyTorch. Our implementation is largely faithful to the description in \cite{o2016convolutional}, though we used the LeakyReLU \cite{maas2013rectifier} activation function instead of ReLU, as we found the LeakyReLU setting had substantially improved trainability. We split the input data of 220,000 labeled examples into 80\% training and 20\% validation sets. Fig.~\ref{fig:Loss_and_Accuracy} shows the average loss and classification accuracy over 100 epochs of training. As expected, the final average accuracy of 53.5\% on the validation set agrees with the result in \cite{o2016convolutional}.

Using the trained weights, we studied a spectrum of performance optimizations for inference. Across all of our benchmarks, we ensure that gradient evaluation was disabled (\texttt{torch.no\_grad()} and Dropout layers are skipped (\texttt{model.eval()}) so that artifacts of training are not counted towards our runtime measurements. All of our inference time benchmarks were run on an AWS EC2 g5.xlarge instance, which has an NVIDIA A10G GPU with 24 GiB of GPU memory, 4 AMD EPYC 7R32 vCPUs, and 16 GiB of memory.

\begin{figure*}[h]
    \centering
    \includegraphics[width=1.0\linewidth]{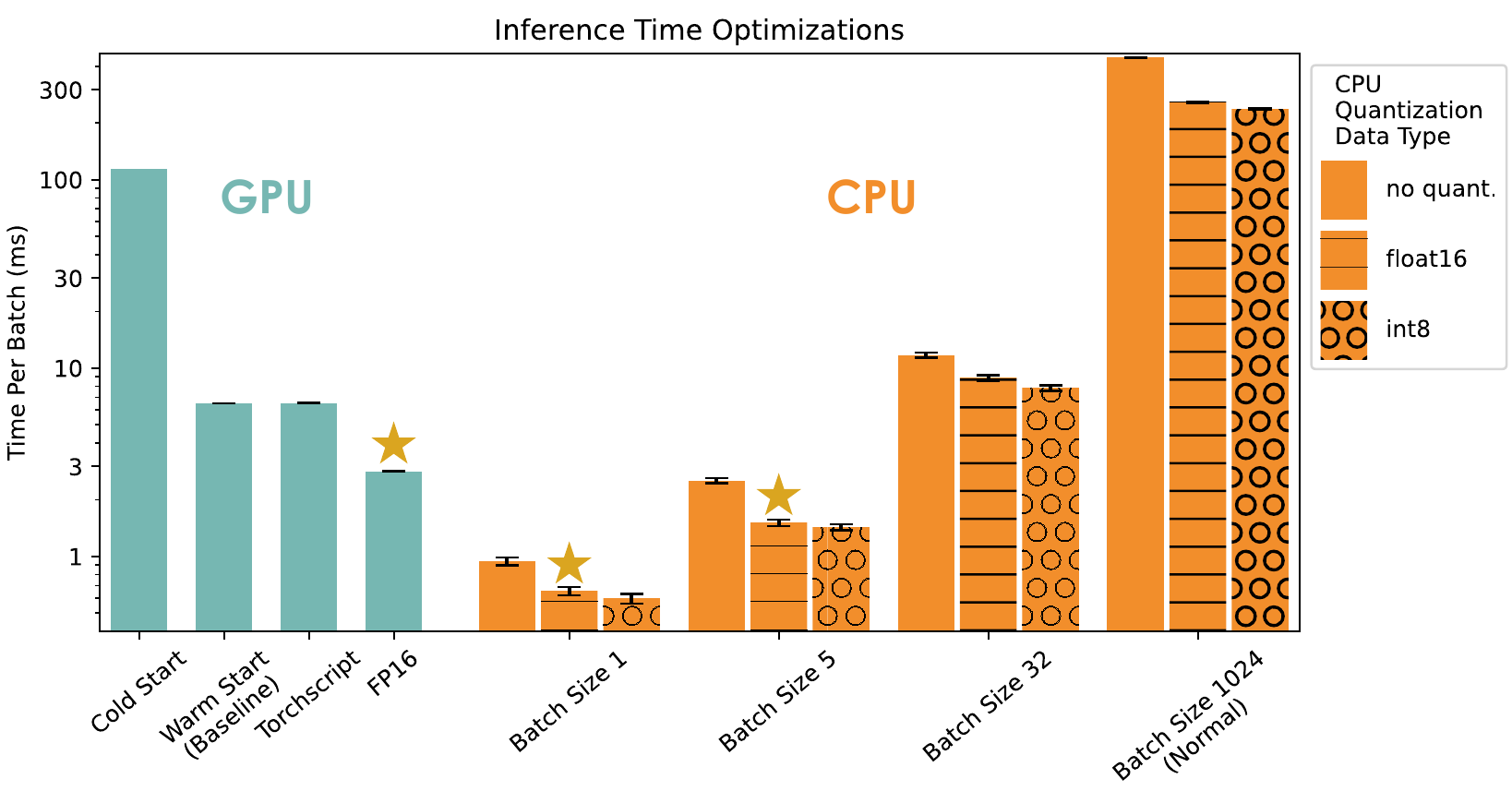}
    \caption{Inference time (note log scale) per batch. For GPU, batch size is always 1024. For CPU, we study batch sizes of 1, 5, 32, and 1024. All implementations are warm-started except for the leftmost Cold Start bar. We highlight the three starred bars as the best choices for real-world deployment, achieving sub-millisecond latency with speedups up to 10$\times$ faster than Warm Start Baseline and 177$\times$ faster than the Cold Start scenario. Error bars represent one standard deviation. \label{fig:inference_time_optimizations}}
\end{figure*}

\subsection{GPU inference}

To establish a ``worst case" latency scenario, we first studied a cold started setting. A cold start refers to the scenario where a model is loaded and initiated on GPU from an inactive state, with no preloaded data or prior computation caching. This situation can occur in real-world applications when GPU hardware restarts or when a model is deployed for the first time without any preliminary execution. While this is an atypical case during continuous operation, it is indeed a case that one would occasionally encounter when serving a model, if one does not take care to explicitly ensure that inference is warmed. The cold start was performed by literally rebooting the GPU through \texttt{nvidia-smi {-}{-}gpu-reset}. For our reference model, we found a cold-start inference time per batch (1024 inputs) of 115 ms. While this is relatively slow, we note that it is faster than the (presumably warm-started) 200 ms per batch inference time achieved in 2016 \cite{o2016radio}; we attribute this largely to GPU hardware improvements.

With this worst case result established, we studied warm starting, whereby we simply ran the model 10 times prior to inference. Warm starting enhances inference speed by keeping the model and its associated data loaded in memory. Prior runs of the model ``warm up" the system, ensuring that data paths are established and computational resources are optimally allocated, reducing latency significantly compared to a cold start. In this warm started setting, we measured inference time of 6.500(2) ms per batch (all errors reported are 1 standard deviation).  This is over $18\times$ faster than the cold start inference time. For further granularity, we separately measured GPU-computation time of 6.329(2) ms and data transfer time between CPU and GPU of 0.187(2) ms, indicating that data transfer is not a bottleneck for latency. Note that these sum to more than 6.500 ms; this is because GPUs can overlap data transfer and compute \cite{harris2012overlap}; but to time separately, we forced a synchronization call that impeded overlapping. We regard the warm started result of 6.5 ms per batch as a Baseline result for typical performance without further optimization. Note that all other scenarios we tested below (except for the cold start described above) also utilized a warm start.

We next used TorchScript \cite{devito2022torchscript} (via \texttt{torch.jit.trace}) which compiles PyTorch models into a form that can be run in a high-performance environment independent of Python. This process optimizes the model for faster execution and potentially better inference performance by enabling integration with lower-level system components and hardware accelerations. However, we found that the traced implementation had a runtime of 6.511(2) ms, which is slightly \textit{worse} than the Baseline warm start setting. At present, we don't fully understand this slowdown; however, we note that other practitioners have also reported similar results on developer forums.

Next, we utilized the Half-precision floating point format (FP16), which reduces the number of bits used to represent numbers in computations from 32 bits (as in single precision) to 16 bits. This reduction is performed simply by calling \texttt{.half()} and can significantly accelerate model inference times by allowing more operations to be performed in parallel and reducing the data transfer load between the CPU and GPU. We found that the FP16 mixed precision GPU implementation had inference time per batch of 2.831(2) ms, which is $2.3\times$ faster than the warm start. A risk of changing precision is degradation in performance in terms of output classification accuracy; however, we found minimal degradation: accuracy on the validation set changed from 53.484\% on the Baseline un-quantized implementation to 53.480\%. This means that the FP16 implementation made just +2 more classification errors on the validation set than the un-quantized implementation did. Just as for the un-quantized baseline, we separately measured the GPU-computation time as 2.647(5) ms and the CPU-GPU data transfer as 0.239(1) ms. This is still dominated by the GPU computations, but is less lopsided than for the baseline. For completeness, Fig.~\ref{fig:fp16_gpu_trace} in the appendix shows a more detailed profiling trace for the FP16 implementation.

\subsection{CPU inference}

We next considered inference for CPUs. As is widely reported in prior literature, while GPU has decisive advantages for training, it is often faster to use CPU to serve trained models for inference. We first tried to run our reference convolutional neural network on CPU with the usual batch size (1024), but found very slow latency of 444(2) ms. As expected, the CPU is quite slow because it is is not optimized for processing large batches.

However, for many RF signal processing applications, it is unlikely that evaluation will require simultaneous inference on batch sizes as large as 1024. We therefore explored batch sizes of 1, 5, and 32 and found inference times of 0.94(5) ms, 2.52(8), and 11.7(4) ms respectively. Particularly for the single-example (batch size 1) inference, this reflects a significant latency reduction over even the FP16 GPU implementation ($3\times$ faster, although recall that the GPU processes a batch size of 1024).

We next explored quantization for CPU inference with int8 and float16 data types. We performed model quantization via PyTorch's \texttt{quantize\_dynamic} function, which is not available for CUDA/GPU. While this model quantization process is similar in spirit to the FP16 half precision implementation used for GPU, it is dynamic in nature, applying scale factors based on computations observed at runtime. While int8 is unsurprisingly the fastest, it has a notable decrease in performance accuracy: for the validation set, accuracy is only 42.143\%, which is a significant degradation in output classification quality. Remarkably however, the float16 quantized implementation has accuracy of 53.482\% which is nearly identical to the unquantized case. Moreover, the float16 CPU model achieves inference times of 0.65(3) ms, 1.50(6) ms, 8.8(3) ms, and 257(2) ms for the four batch sizes. Again, at the large batch sizes, the CPU implementation is slow, but the settings with batch size 1 and 5 achieve very low latency with virtually no degradation in validation accuracy. For completeness, Fig.~\ref{fig:float16_cpu_trace} in the appendix shows a more detailed profiling trace for the float16 dynamically quantized implementation.

\subsection{Discussion}

Fig.~\ref{fig:inference_time_optimizations} presents all of our inference time results. Overall, we conclude that the three starred implementations are best-suited for real world deployment: FP16 half-precision on GPU batch size 1024 (2.83 ms), float16 dynamic quantization on CPU with batch size 1 (0.65 ms), and float16 dynamic quantization on CPU with batch size 5 (1.50 ms). While all three starred implementation involve quantization, output classification accuracy is nearly unchanged for these implementations. However, quantization must be applied carefully: while the int8 dynamically quantized CPU settings unsurprisingly achieve the fastest inference times, there is a steep decline in classification accuracy.

These three implementations are 2.3--10$\times$ faster than the Baseline warm start GPU setting and 41--177$\times$ faster than the worst-case cold start GPU setting. The choice between the three implementations is a matter of whether batch inference is necessary. If not, CPU is well-suited for real-time processing. Even if batch inference is necessary, at small scales (like batch size 5), CPU is still be preferable. However, at large batch size settings (certainly including training), GPU has a significant advantage. Again, while our results are benchmarked on the CNN2 reference convolutional neural network, the key takeaways are not exclusive to this model and generalize to many deep learning models.

\section{Quantum RF Signal Processing} \label{sec:quantum_rf_signal_processing}
\begin{figure}[h]
    \centering    
    \includegraphics[width=0.5\linewidth]{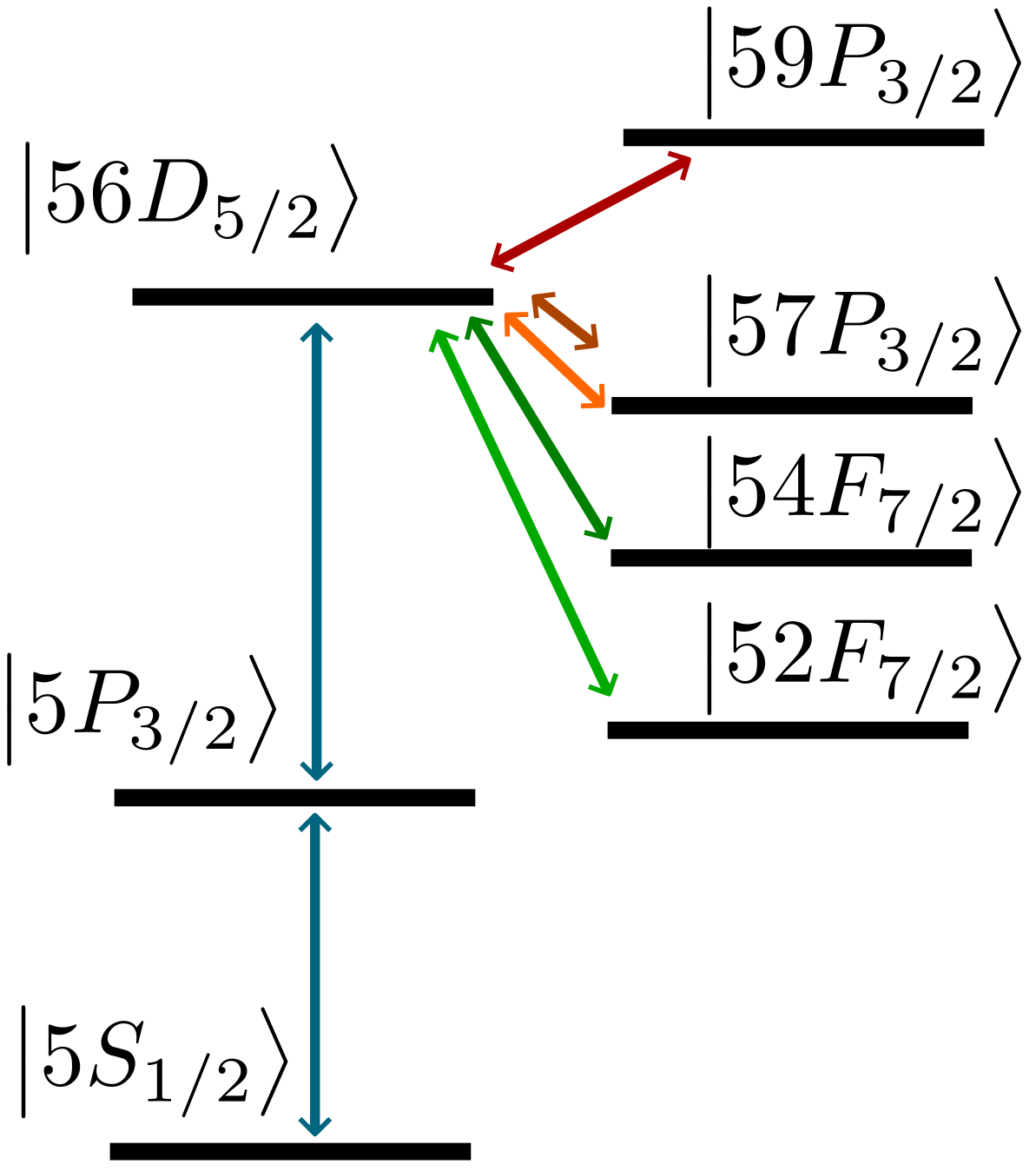}
    \caption{Atomic level diagram reproduced from Fig. 3a of Miller et al. \cite{miller2023rydiqule}, atomic level diagram. The five colored arrows (red, brown, orange, two green) on the right indicate the five 1.72, 12.1, 27.4, 65.1, and 116 GHZ tones that are sensed, via four Rydberg transitions.\label{fig:rydiqule_transitions_creativecommons}}
\end{figure}
We finally turn our attention to the portability of deep learning methods for RF classification to emerging quantum RF hardware. As described in Section~\ref{sec:introduction}, QRF sensors based on Rydberg atoms have multiple advantages over traditional RF sensors including broad frequency range, high resolution, high sensitivity, and minimal disturbance to sensed signals. Recent hardware progress has advanced QRF technology closer to real-world deployment. For instance, a recent test of Infleqtion's SqyWire QRF receiver at NetModX23 demonstrated ultra-wideband coverage spanning spectrum from HF (3--30 MHz) to SHF (3--30 GHz) \cite{infleqtions2023quantum}.

Ideally, progress in AI/ML techniques for RF processing should transfer gracefully to the QRF setting. 
Prior work has already demonstrated the potential of applying classical AI/ML techniques to enhance both quantum computing and sensing technologies. For instance, regression models have been used to reduce quantum error mitigation overheads \cite{liao2023machine} and generate noise-resilient circuits \cite{cincio2021machine}. Additionally, reinforcement learning approaches have been developed that generate more robust quantum sensing protocols with improved signal-to-noise ratios \cite{xiao2022parameter, alam2024robust, belliardo2024application}.
QRF receivers present particularly interesting opportunities for deeper integration with AI/ML models to efficiently and accurately analyze the data output by the quantum sensor. Note in particular that a single physically compact broadband (Hz to THz) QRF receiver could replace multiple state-of-the-art conventional RF receivers, leading to significant reductions in size, weight, power, and cost. However, such a device would also need robust software support to collectively process a wide band of frequencies (whereas previously each conventional RF receiver could typically have a separate processor).

Our focus here is on a proof of concept demonstrating portability of our deep neural networks for RF classification from conventional RF to QRF datasets. We leverage RydIQule (Rydberg Interactive Quantum module) \cite{miller2023rydiqule}, an open-source Python library that simulates the interaction of Rydberg atoms with RF inputs. RydIQule leverages the Alkali Rydberg Calculator (ARC) software package \cite{vsibalic2017arc} to insulate end users from needing to specify properties of Rydberg atoms, enabling a higher programming abstraction.

Our proof of concept is based on RydIQule's marquee simulation of a recent experiment \cite{meyer2023simultaneous} in which a single QRF sensor based on a Rubidium vapor cell was used to simultaneously detect and demodulate signals from five RF tones (frequencies) spanning from 1.72 GHZ to 116 GHZ. This task would otherwise require multiple receivers with separate antennas. Fig.~\ref{fig:rydiqule_transitions_creativecommons} shows the level diagram for this multi-band demodulation experiment (and corresponding simulation). We refer readers to \cite{meyer2023simultaneous, vsibalic2017arc} for details of the physical models and the underlying electromagnetically induced transparency (EIT) and Autler-Townes phenomena that enable sensing.

\begin{figure*}[h]
    \centering    \includegraphics[width=0.95\linewidth]{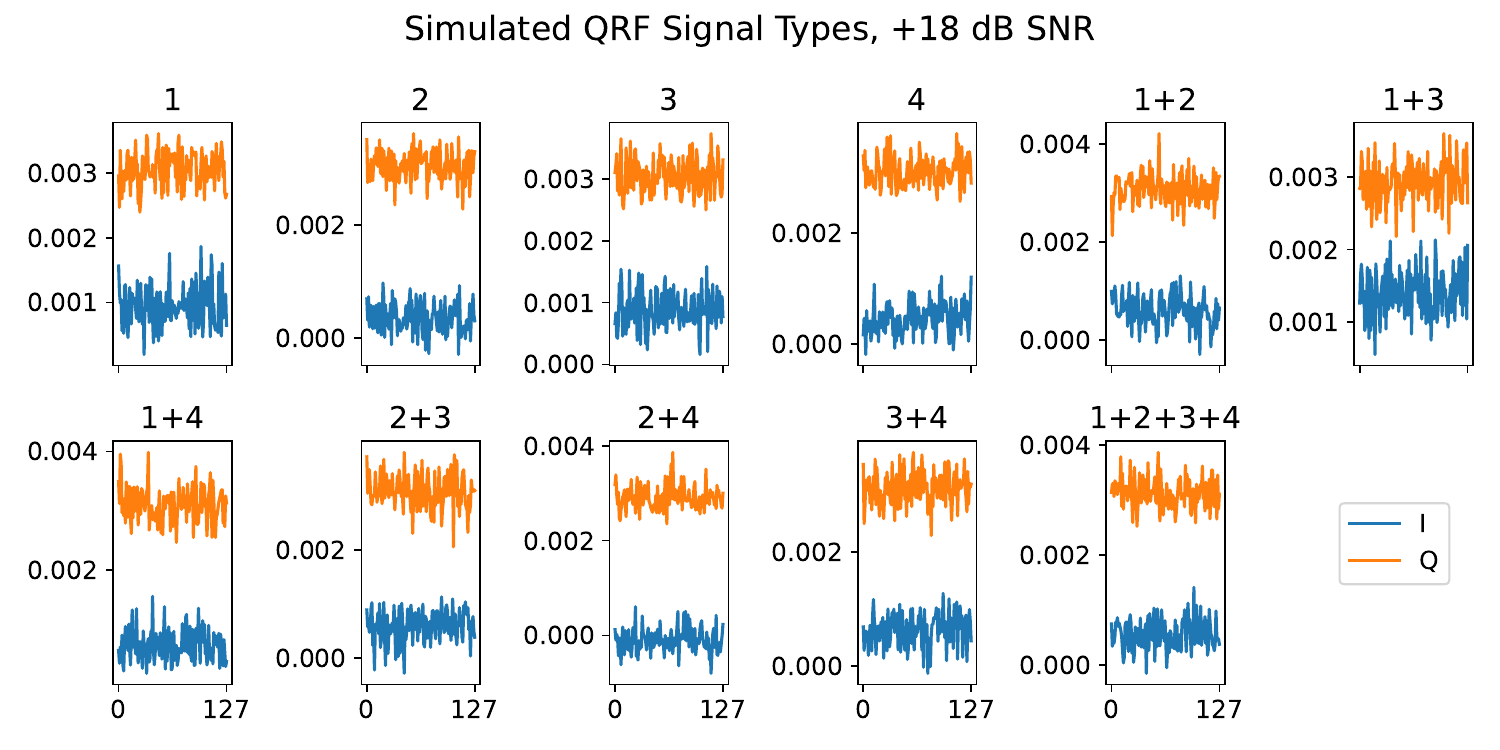}
    \caption{Examples of the 11 QRF simulated outputs, in time domain. Tone 1, 2, 3, 4, and 5 correspond to frequencies 1.72, 12.1, 27.4, 65.1, and 116 GHZ.} 
    \label{fig:QRF_Signal_Types}
\end{figure*}

For our purposes, we regard RydIQule as a black box that simulates the response of the QRF sensor to different RF tones and results in a timeseries output signal. We used RydIQule to simulate outputs for 11 linear combinations of the five input tones. To draw a parallel with the \texttt{RadioML2016.10A} dataset studied in Section~\ref{sec:online_classification}, we generated 1000 samples across 20 SNRs (-20 to +18 dB) for each of the 11 signal types, for a total of 220,000 labeled examples. High-SNR examples of these output signals are shown in Fig.~\ref{fig:QRF_Signal_Types}. Unlike the modulations in the \texttt{RadioML2016.10A} dataset, these 11 signals cannot be easily visually distinguished (cf. Fig.~1 in \cite{o2016convolutional}) in the time domain.

\begin{figure*}[h]
    \centering    \includegraphics[width=0.8\linewidth]{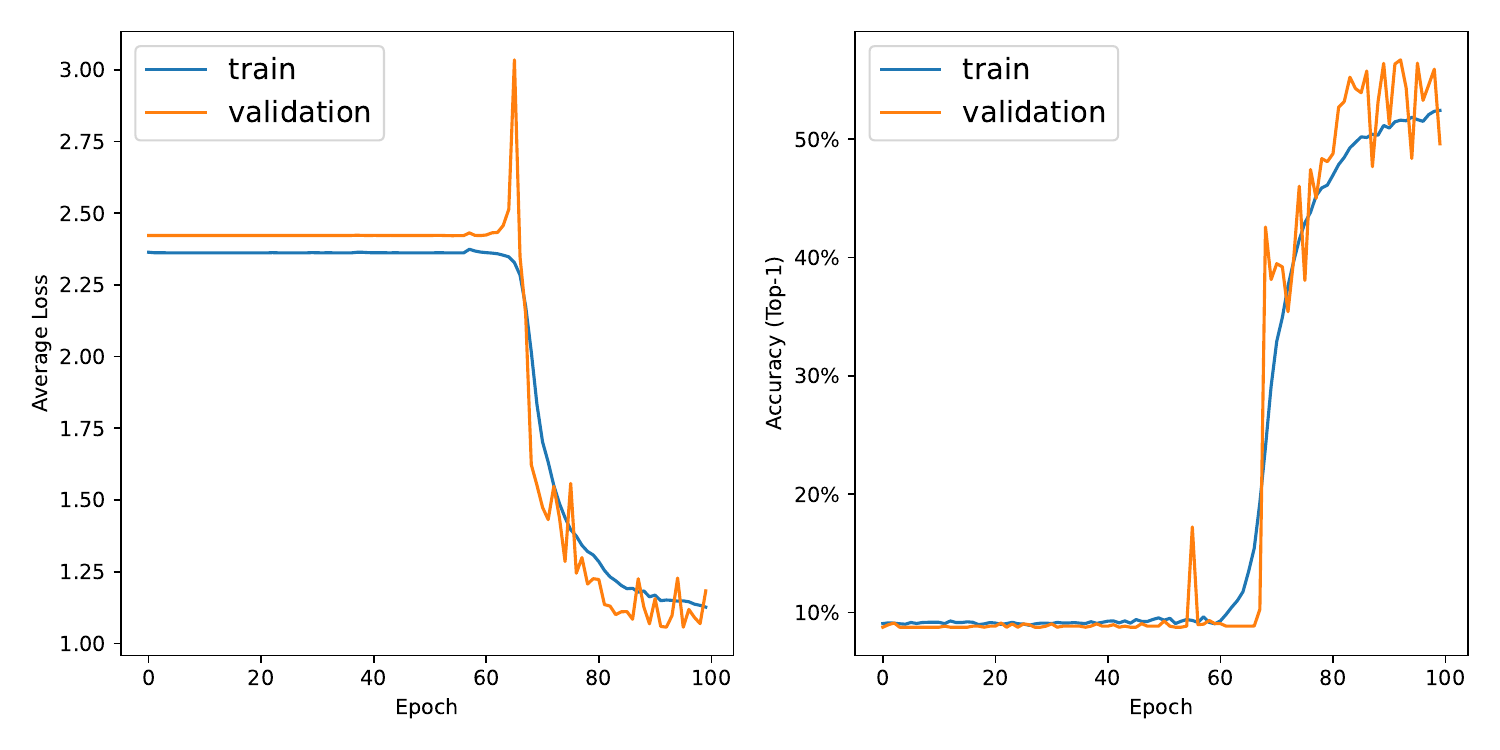}
    \caption{Loss (left) and classification accuracy on the our RydIQule-simulated QRF sensing dataset using the same CNN2 deep convolutional neural network applied in Section~\ref{sec:latency_optimization}.  After 100 epochs of training, the model achieves a classification accuracy of 56.4\%.}
    \label{fig:QRF_Loss_and_Accuracy}
\end{figure*}

We applied the deep convolutional neural network from Section~\ref{sec:latency_optimization} to this dataset of QRF outputs. The results are shown in Fig.~\ref{fig:QRF_Loss_and_Accuracy}. While the underlying dataset is more challenging to classify than \texttt{RadioML2016.10a}, the trained model encouragingly achieves a classification accuracy of 56.4\% on the validation set.

\begin{figure}
  \centering
    \includegraphics[width=0.5\textwidth]{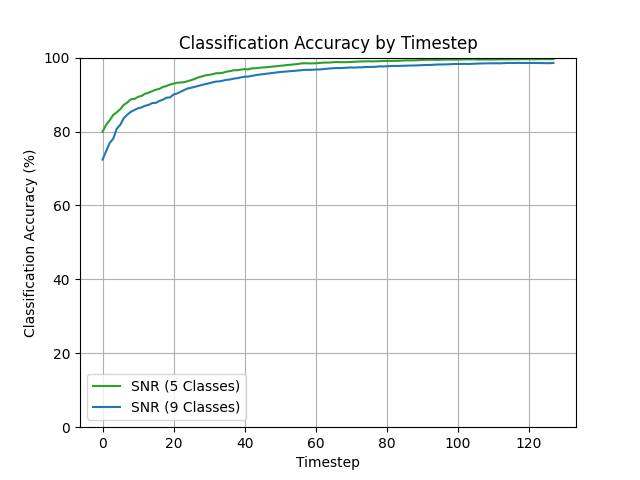}
    \caption{Top-1 classification accuracy as a function of timestep for the CWT-RNN model on the 5-SNR and 9-SNR classification tasks over the QRF validation dataset. Both tasks ultimately achieve $> 98\%$ accuracy, with 
    $> 70\%$ accuracy achieved from the first timestep.\label{fig:qrf_snr_classification}}
\end{figure}

Furthermore, we applied our novel CWT-RNN model described in Section~\ref{sec:online_classification} to the QRF dataset and evaluated performance on the task of SNR classification. The online average classification accuracy as a function of timesteps ingested is shown in Fig.~\ref{fig:qrf_snr_classification} for the 5-SNR and 9-SNR classification tasks. Training and validation details are largely the same as presented in Section~\ref{sec:online_classification} for the \texttt{RadioML2016.10a} dataset. Remarkably, the CWT-RNN approach exhibits improved performance on the QRF dataset, with both the 5-SNR and 9-SNR classification tasks obtaining top-1 classification accuracies above 98\%, with $> 70\%$ accuracy from the first timestep. These results indicate that the suitability of our approach extends to the domain of QRF signals and that CWT-RNN models may be used to enable real-time signal understanding in quantum RF receivers. 

We caution that our simulated dataset does not necessarily reflect all design details in real QRF systems, on which we expect significantly better performance than possible with traditional RF systems. Again, our primary aim here is simply to validate portability of deep learning methods for traditional RF sensors to quantum RF sensors.

\section{Conclusion} \label{sec:conclusion}

We summarize our key contributions as follows:
\begin{itemize}
    \item Development of deep learning models that enable \textit{online} classification of RF signals on-the-fly by leveraging Continuous Wavelet Transforms (CWT) and Recurrent Neural Networks (RNN). This approach enables lower-latency RF classification than previous AI/ML approaches, which generally issue a reliable classification prediction only once the entire input timeseries has been processed.
    \item Inference optimizations across both CPU and GPU that enable sub-millisecond prediction latency---with over two orders of magnitude of latency improvement spanned. Our optimizations are benchmarked for a specific deep convolutional neural network, but apply more generally across many deep learning models.
    \item Validated portability of our deep learning approach to emerging QRF sensing hardware that exhibit several improvements over traditional RF sensors including broad frequency range, high resolution, high sensitivity, and minimal disturbance to sensed signals. We leveraged the RydIQule physics simulation library to generate a realistic dataset of QRF sensor outputs under different input RF tones.
\end{itemize}

Altogether, our work bridges towards next-generation RF sensors that use quantum technology to surpass previous physical limits, paired with latency-optimized AI/ML software that is suitable for real-time deployment

While our results are promising in terms of latencies required for radar applications, lower latencies could be achieved to enable other applications with more stringent requirements.
In particular, hardware synthesis tools \cite{soda} could enable ASIC (Application-Specific Intergrated Circuit) implementations with orders-of-magnitude lower latency and/or energy-efficiency.  Furthermore, these approaches could be used in concert with more aggressive mixed-precision techniques \cite{Wang_2019_CVPR} to further improve performance.

In addition, while our proof of concept leveraged RydIQule simulation of QRF sensing in a black box setting, we suspect that cross-layer optimization that integrates deeper into the physics stack will be profitable. Similar efforts with our Superstaq software platform \cite{campbell2023superstaq}, have been able to improve utilization of quantum computing hardware through deeper integration with underlying device physics.

\section*{Acknowledgment}
We thank Corey Heitzman and Seth Caliga of Infleqtion for their technical input and guidance.

\bibliographystyle{unsrt}
\bibliography{refs}

\section*{Appendix: Profiling Traces}
(Next page)
\begin{figure*}[h]
    \centering
    \includegraphics[width=0.99\linewidth]{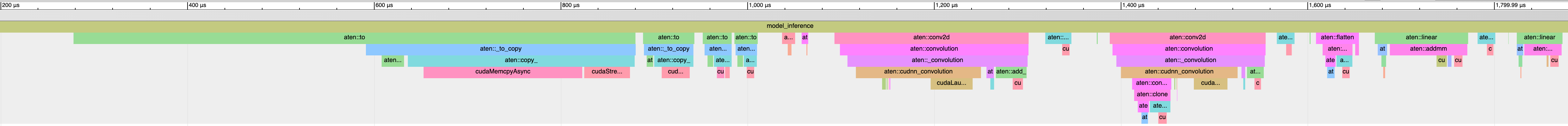}
    \caption{Execution trace for inference on FP16 GPU. Compute latency is dominated by the 2D convolution.}
    \label{fig:fp16_gpu_trace}
\end{figure*}

\begin{figure*}[h]
    \centering
    \includegraphics[width=1.0\linewidth]{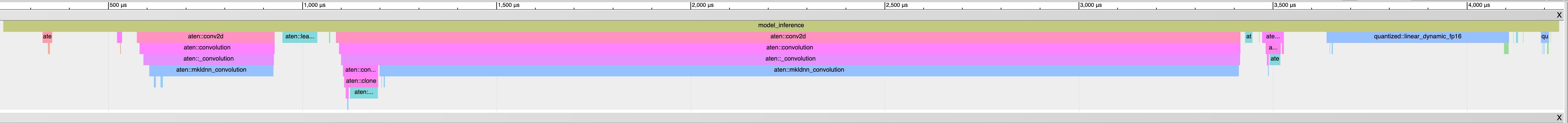}
    \caption{Execution trace for inference on float16 CPU. Latency is once again dominated by the 2D convolution. Note that on CPU, there no data transfer required and therefore no \texttt{aten::to} call.}
    \label{fig:float16_cpu_trace}
\end{figure*}

\end{document}